\documentclass[11pt]{article}
\usepackage{amsmath}
\usepackage{amssymb}
\usepackage[dvips]{epsfig}

 \advance \textwidth by -2in
 \advance \textheight by -3in
\def\##1{\underline #1}
\def\=#1{\underline{\underline #1}}

\def\eps{\epsilon}

\def\ko{k_0}
\def\co{c_0}

\def\.{\mbox{ \tiny{$^\bullet$} }}

\def\epsa{\epsilon_a}
\def\epsb{\epsilon_b}
\def\mua{\mu_a}
\def\mub{\mu_b}

\def\ux{\#{u}_x}

\def\uz{\#{u}_z}
\def\uc{\#{u}_c}
\def\ua{\#{u}_k}

\def\le{\left(}
\def\ri{\right)}
\def\les{\left[}
\def\ris{\right]}
\def\lec{\left\{}
\def\ric{\right\}}

\def\c#1{\cite{#1}}

\def\r#1{(\ref{#1})}

\begin{document}
%\noindent Submitted for publication in {\it Optik}
%\vskip 0.4cm

\vskip 0.4cm

\begin{center}
{\large {\bf Counterposed phase velocity and energy--transport velocity vectors in a dielectric--magnetic uniaxial medium}}
\vskip 0.2cm

\noindent  {Akhlesh Lakhtakia}\footnote{Corresponding Author. Tel: +1 814 863 4319; Fax: +1 814 865 9974;
E--mail: AXL4@psu.edu}
\vskip 0.2cm
\noindent {\em Computational \& Theoretical Materials Sciences Group (CATMAS)\\
Department of Engineering Science \& Mechanics\\
Pennsylvania State University, University Park, PA 16802--6812, USA}
\vskip 0.2cm
\noindent  {Martin W. McCall}
\vskip 0.2cm
\noindent {\em Department of Physics, Imperial College London\\
Prince Consort Road, London SW7 2BZ, UK}

\end{center}

\noindent {\bf Abstract:}
When a plane wave is launched from a plane surface in a linear, homogenous, dielectric--magnetic,
uniaxial medium, we show that its phase velocity and the energy--transport velocity vectors  can be counterposed (i.e.,
lie on different sides
of the surface normal) under certain circumstances.

\vskip 0.2cm
\noindent {\em Keywords:\/} Anisotropy; Energy--transport velocity; Phase velocity;

\section{Introduction}
In any linear homogeneous medium,  two distinct plane waves can propagate in any direction (except in very
rare circumstances \c{Ger,Ber}, which are ignored here). With each plane wave are associated a phase velocity
vector and an energy--transport velocity vector \c{Chen}. These two vectors are parallel to each other in isotropic
mediums, but not in anisotropic mediums. 

\begin{figure}[!ht]
\centering \psfull
\epsfig{file=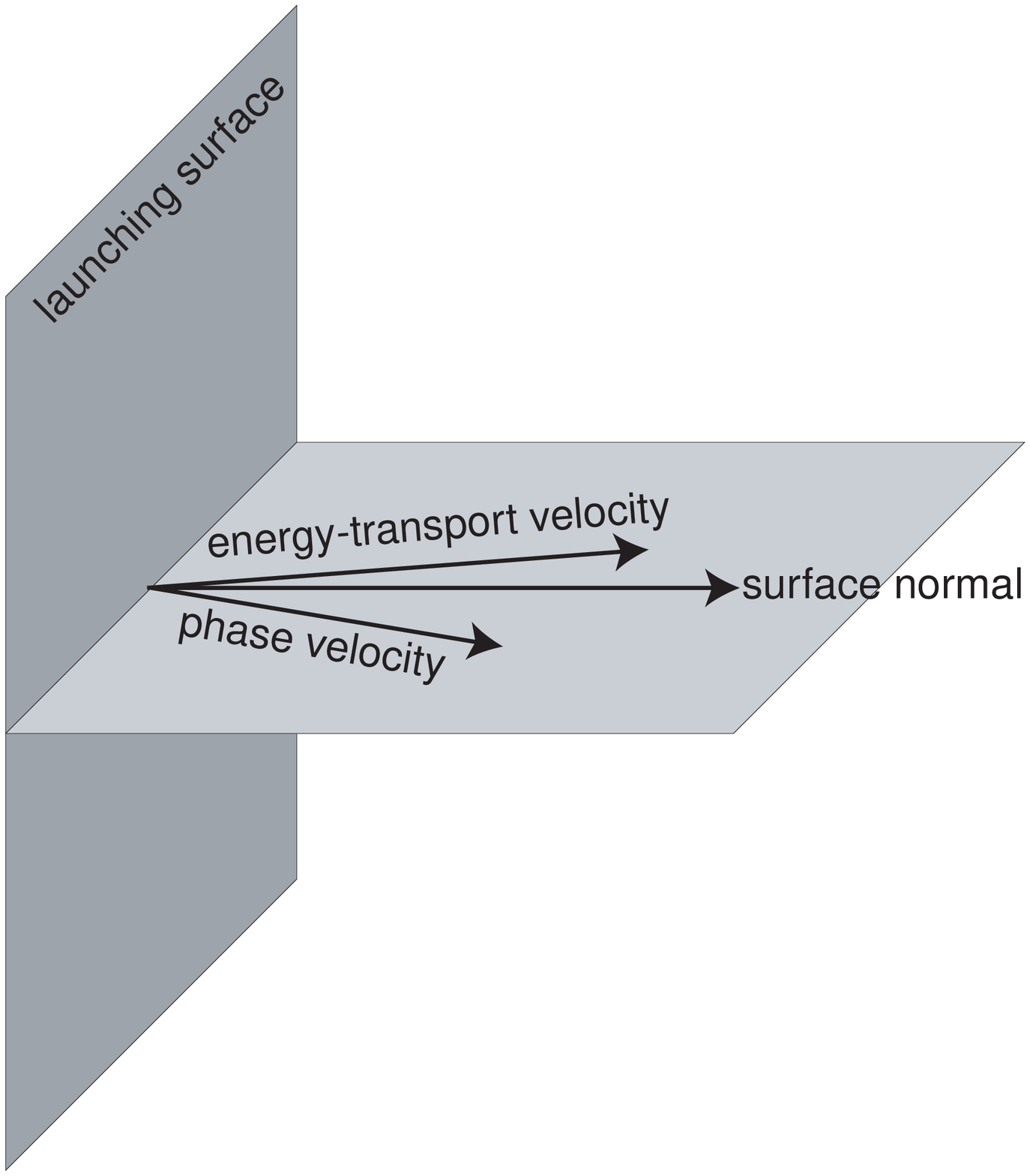,width=8cm }
\caption{Schematic to explain counterposition of the phase velocity and the energy transport
velocity vectors.
}
\label{Fig1}
\end{figure}

While examining  a recently reported experimental result \c{ZFM}, we
came across the following question: If a plane wave is launched from an infinite plane~---~possibly,
either by reflection or refraction~---~into a linear, homogeneous,
anisotropic medium, can the phase velocity and the energy--transport velocity vectors be counterposed (i.e.,
lie on different sides
of the surface normal), as shown in Figure
\ref{Fig1}? Although we suspected an affirmative answer to the question, we were unable
to find any treatment of the question in standard textbooks. Therefore, we undertook an investigation, the results of which are
reported here. 

\section{Analysis}
Let us consider a dielectric--magnetic uniaxial medium whose relative permittivity and relative permeability
dyadics are denoted by
\begin{eqnarray}
&&\=\eps_r = \epsa\, \=I + (\epsb-\epsa) \uc\uc\,,\\
&&\=\mu_r = \mua\,\=I + (\mub-\mua)\uc\uc\,,
\end{eqnarray}
respectively, where $\=I$ is the identity dyadic and $\uc$ is a unit vector parallel to the
distinguished axis of the medium.

In this medium, two distinct  plane waves can propagate in any given direction,
as detailed elsewhere \c{LVV91}. 
The wavenumbers of the two plane waves are obtained as
\begin{eqnarray}
&& k_1 = \ko \le\frac{\mua\epsa\epsb}{\ua\.\=\eps_r\.\ua}\ri^{1/2}\,,\\[8pt]
&& k_2 = \ko \le\frac{\epsa\mua\mub}{\ua\.\=\mu_r\.\ua}\ri^{1/2}\,,
\end{eqnarray}
where $\ua$ is a unit vector denoting the direction of
propagation while $\ko$ is the free--space
wavenumber. The electric and magnetic field phasors associated with the two
plane waves are known, their expressions not being needed for the present
purposes.

But we do need expressions for the phase velocity and the energy--transport velocity vectors. With the assumption
that the imaginary parts of $\eps_{a,b}$ and $\mu_{a,b}$ are negligibly small, we obtain \c{LVV91}
\begin{equation}
{\#v}_{p_\ell} = \frac{\omega}{k_\ell}\, \ua\,,\quad (\ell=1,2)\,,
\end{equation}
for the phase velocity vectors, and
\begin{equation}
\left.\begin{array}{ll}
{\#v}_{e_1} = \co\,\frac{k_1}{\ko}\, \frac{\=\eps_r\.\ua}{\,\,\mua\epsa\epsb\,\,}\\[8pt]
{\#v}_{e_2} =  \co\,\frac{k_2}{\ko}\, \frac{\=\mu_r\.\ua}{\,\,\epsa\mua\mub\,\,}
\end{array}\right\}
\label{blah}
\end{equation}
for the energy--transport velocity vectors of the two plane waves, with $\co$ denoting the
speed of light in free space. Note that ${\#v}_{e_{1,2}}$
are co--parallel with the respective time--averaged Poynting vectors; and they are also
co--parallel with the respective group velocity vectors in the absence of dispersion \cite[Sec. 3.6]{Chen}.

Let us now suppose that a plane wave is launched into the half--space $z>0$ from the
plane $z=0$. We say that the phase velocity vector ${\#v}_{p_\ell}$ and
the energy--transport velocity vector ${\#v}_{e_\ell}$  
of the $\ell$-th plane wave, ($\ell =1,2$),
are {\em  counterposed}  if the two vectors are pointed
on the opposite sides of the $+z$ axis.

Without loss of generality, we set
\begin{equation}
\left.\begin{array}{ll}\uc = \sin\xi\,\ux + \cos\xi\,\uz\\
\ua = \sin\theta\,\ux+\cos\theta\,\uz
\end{array}\right\}\,,
\end{equation}
where $\ux$ and $\uz$ are unit cartesian vectors, while the angles
$\theta\in\les -90^\circ,90^\circ\ris$ and $\xi\in\les -90^\circ,90^\circ\ris$. Then, the expressions 
\begin{eqnarray}
\nonumber
{\#v}_{e_1} &=&  \frac{k_1}{\ko}\, \frac{\co}{2\mua\epsa\epsb}\,
\lec
(\epsa+\epsb)(\sin\theta\ux+\cos\theta\uz)\right.\\
&&\qquad+\left.
(\epsa-\epsb)\les\sin(\theta-2\xi)\ux -\cos(\theta-2\xi)\uz\ris\ric
\end{eqnarray}
and
\begin{eqnarray}
\nonumber
{\#v}_{e_2} &=&  \frac{k_1}{\ko}\, \frac{\co}{2\epsa\mua\mub}\,
\lec
(\mua+\mub)(\sin\theta\ux+\cos\theta\uz)\right.\\
&&\qquad+\left.
(\mua-\mub)\les\sin(\theta-2\xi)\ux -\cos(\theta-2\xi)\uz\ris\ric\,
\end{eqnarray}
emerge from \r{blah}. 

Let us define angles $\psi_{\ell}$, ($\ell=1,2$), through the relation ${\#v}_{e_{\ell}} =
{v}_{e_{1,2}} (\sin\psi_\ell\ux+\cos\psi_\ell\uz)$; hence,
\begin{equation}
\tan\psi_\ell=\frac{\sin\theta - \delta_\ell\, \sin(\theta-2\xi)}{\cos\theta + \delta_\ell\, \cos(\theta-2\xi)}\,,\quad (\ell=1,2)\,,
\end{equation}
where
the degree of uniaxiality
\begin{equation}
\delta_\ell =\lec\begin{array}{ll}
 \frac{\epsb-\epsa}{\epsb+\epsa}\,,&\qquad {\rm if} \,\,\ell=1\\[8pt]
\frac{\mub-\mua}{\mub+\mua}\,,&\qquad{\rm if} \,\, \ell=2
\end{array}\right.\,\,.
\end{equation}
The counterposition condition
then amounts to
\begin{equation}
(\sin\theta)\,(\tan\psi_\ell) < 0\,.
\end{equation}
Alternatively, the two velocity vectors of the
 $\ell$-th plane wave  are counterposed if $\psi_{\ell} \lessgtr 0^\circ$ when
$\theta \gtrless 0^\circ$.

\begin{figure}[!ht]
\centering \psfull
\epsfig{file=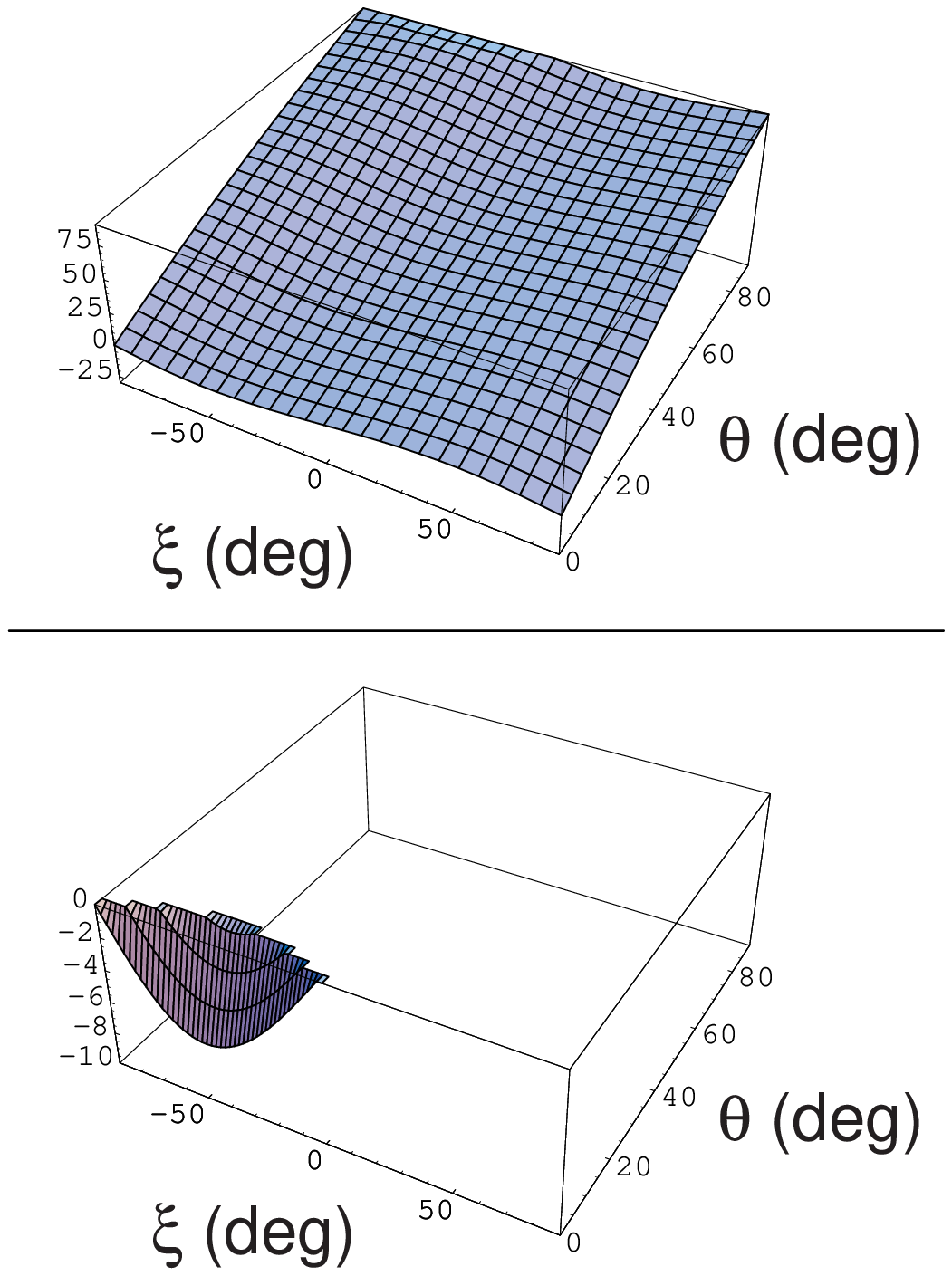,width=8cm }
\caption{Top: Computed values of $\psi_\ell$ for $\theta \in\les 0^\circ,90^\circ\ris$, when $\delta_\ell=1/9$. 
Bottom: Negative values of
$\psi_\ell$ are isolated in this graph.
}
\label{Fig2}
\end{figure}

Figure \ref{Fig2}  shows computed values of $\psi_{\ell}$ for $\theta > 0^\circ$, when the degree of uniaxiality
is positive (i.e., $\epsb>\epsa$ for $\ell = 1$, and $\mub>\mua$ for $\ell = 2$). Figure 3
shows the computed values for negative uniaxiality  (i.e., $\epsb<\epsa$ for $\ell = 1$, and $\mub<\mua$ for $\ell = 2$).
The latter figure can, in fact, be deduced from Figure \ref{Fig2} {\em via} the substitution
$\lec\delta_\ell\to-\delta_\ell,\,\xi\to\xi\pm\pi/2\ric$, but has been included for completeness.
Quite clearly, a wide $\xi$--range exists for very small angles $\theta$ for which the counterposition
condition is satisfied. As $\theta$ increases, the $\xi$--range for counterposition diminishes and eventually vanishes.
The higher the degree of uniaxiality in magnitude, the larger is the portion of the $\theta\xi$ space wherein the counterposition
condition is satisfied.

\section{Conclusion}
When a plane wave is launched from a plane surface~---~possibly by refraction or
reflection~---~in a linear, homogenous, dielectric--magnetic,
uniaxial medium, we have shown
here that its phase velocity and the energy--transport velocity vectors  {\em may} be counterposed. An excellent
experimental example has been furnished by Zhang {\em et al.} \c{ZFM}. 

%\newpage

\newpage

\newpage
\begin{figure}[!ht]
\centering \psfull
\epsfig{file=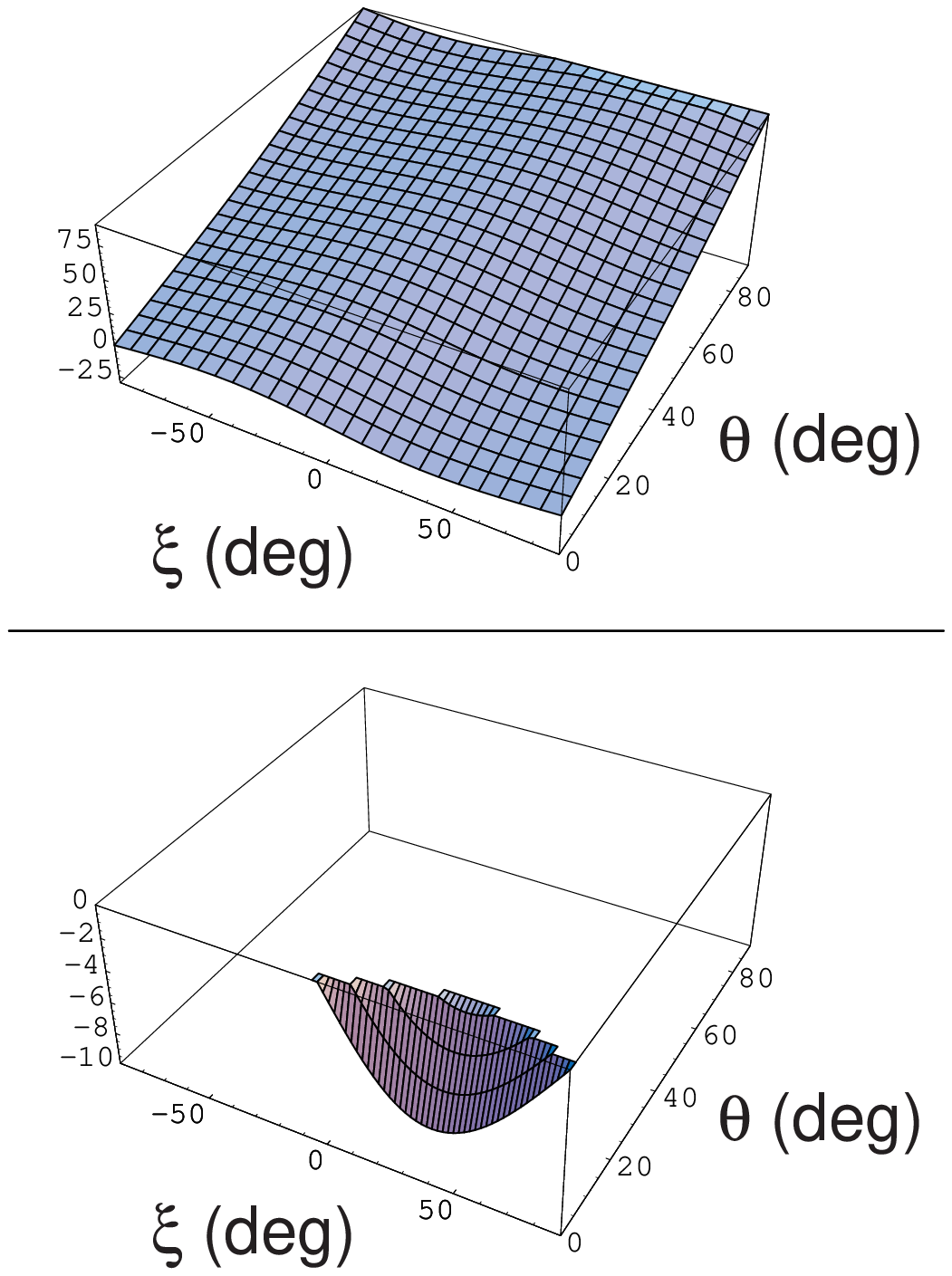,width=8cm }
\caption{Same as Figure \ref{Fig2}, but for $\delta_\ell=-1/9$.
}
\label{Fig3}
\end{figure}

\end{document}